\documentclass[12pt, draftcls, onecolumn]{IEEEtran}

\usepackage{graphicx}
\usepackage{subfigure}
\usepackage{epstopdf}
\usepackage{epsfig}
\usepackage[cmex10]{amsmath}
\usepackage{amssymb}
\usepackage{times}
\usepackage{ntheorem}
\usepackage{pifont}
\usepackage{stfloats}
\usepackage{bm}
\usepackage{color}
\usepackage{makecell}
\usepackage{array}
\usepackage{algorithm}
\usepackage{algorithmic}
\usepackage{multirow}
\usepackage{epsfig,latexsym}
\usepackage{flushend}
\usepackage{cite}
\usepackage{verbatim}
\usepackage{amsopn}
\usepackage{booktabs}
\usepackage{hyperref}

\begin{document}
%
\title{Acquisition of Channel State Information for mmWave Massive MIMO: Traditional and Machine Learning-based Approaches}
\author{Chenhao~Qi,~\IEEEmembership{Senior~Member,~IEEE}, Peihao~Dong,~\IEEEmembership{Member,~IEEE}, Wenyan Ma,~\IEEEmembership{Student~Member,~IEEE},\\ Hua Zhang,~\IEEEmembership{Member,~IEEE}, Zaichen~Zhang,~\IEEEmembership{Senior~Member,~IEEE} and Geoffrey Ye Li,~\IEEEmembership{Fellow,~IEEE}
\thanks{Chenhao Qi, Wenyan Ma, Hua Zhang and Zaichen Zhang are with the School of Information Science and Engineering, Southeast University, Nanjing 210096, China (Email: \{qch,huazhang, 220170772, zczhang\}@seu.edu.cn).}
\thanks{Peihao Dong is with the College of Electronic and Information Engineering, Nanjing University of Aeronautics and Astronautics, Nanjing 211106, China (e-mail: phdong@nuaa.edu.cn).}
\thanks{Geoffrey Ye Li is with the Department of Electrical and Electronic Engineering, Imperial College London, London SW7 2AZ, UK (Email: geoffrey.li@imperial.ac.uk).}
}

\IEEEtitleabstractindextext{%
\begin{abstract}
The accuracy of channel state information (CSI) acquisition directly affects the performance of millimeter wave (mmWave) communications. In this article, we provide an overview on CSI acquisition, including beam training and channel estimation for mmWave massive multiple-input multiple-output systems. The beam training can avoid the estimation of a high-dimension channel matrix while the channel estimation can flexibly exploit advanced signal processing techniques. In addition to introducing the traditional and machine learning-based approaches in this article, we also compare different approaches in terms of spectral efficiency, computational complexity, and overhead.
\end{abstract}

}

\maketitle

\IEEEdisplaynontitleabstractindextext

%
\IEEEpeerreviewmaketitle

\section{Introduction}

Millimeter wave (mmWave) communications have attracted extensive interests from the academia, industry, and government as it can make full use of abundant frequency resources at high frequency band to achieve ultra-high-speed data transmission~\cite{heath2016overview,JSACmmWaveMXiao2017,CM2019HardwareConstrainedmmWave,WC2015mmWaveJYGuo}. The mmWave communication systems are usually equipped with large antenna arrays, known as mmWave massive multiple-input multiple-output (MIMO), to generate highly directional beams and compensate for the severe path loss in high frequency band. However, the performance of directional beamforming largely relies on the accuracy of channel state information (CSI) acquisition. Compared to the traditional MIMO systems, the CSI acquisition in mmWave massive MIMO systems is challenging. On one hand, the large antenna arrays form a high-dimension channel matrix, whose estimation consumes more resources, e.g., pilot sequence overhead, sounding beam overhead, and computational complexity. On the other hand, the mmWave massive MIMO typically employs a hybrid beamforming architecture, where the radio frequency (RF) chains are much fewer than the antennas. Therefore, we can only obtain a low-dimension signal from the RF chains instead of directly getting a high-dimension signal from the frontend antennas, which makes CSI acquisition much more challenging than usual.

CSI acquisition includes beam training and channel estimation. The beam training sounds the mmWave massive MIMO channel with analog transmit and receive beams to find the beam pairs best fitting for the transmission, which can avoid the estimation of a high-dimension channel matrix. The channel estimation focuses on estimating a high-dimension channel matrix, which flexibly exploits advanced signal processing techniques, such as compressed sensing (CS). Both beam training and channel estimation can exploit machine learning (ML) techniques in addition to the traditional approaches.

In this article, we provide an overview on CSI acquisition for mmWave massive MIMO. We first discuss beam training approaches, including beam sweeping, hierarchical beam training, and ML-based beam training. Then we present channel estimation approaches, including the CS-based sparse channel estimation, array signal processing based channel estimation, and ML-based channel estimation. Finally, we compare different approaches in terms of spectral efficiency (SE), computational complexity, and incurred overhead, and identify some future research topics in this area.

\begin{figure*}[t]
\centering
\includegraphics[width=6 in]{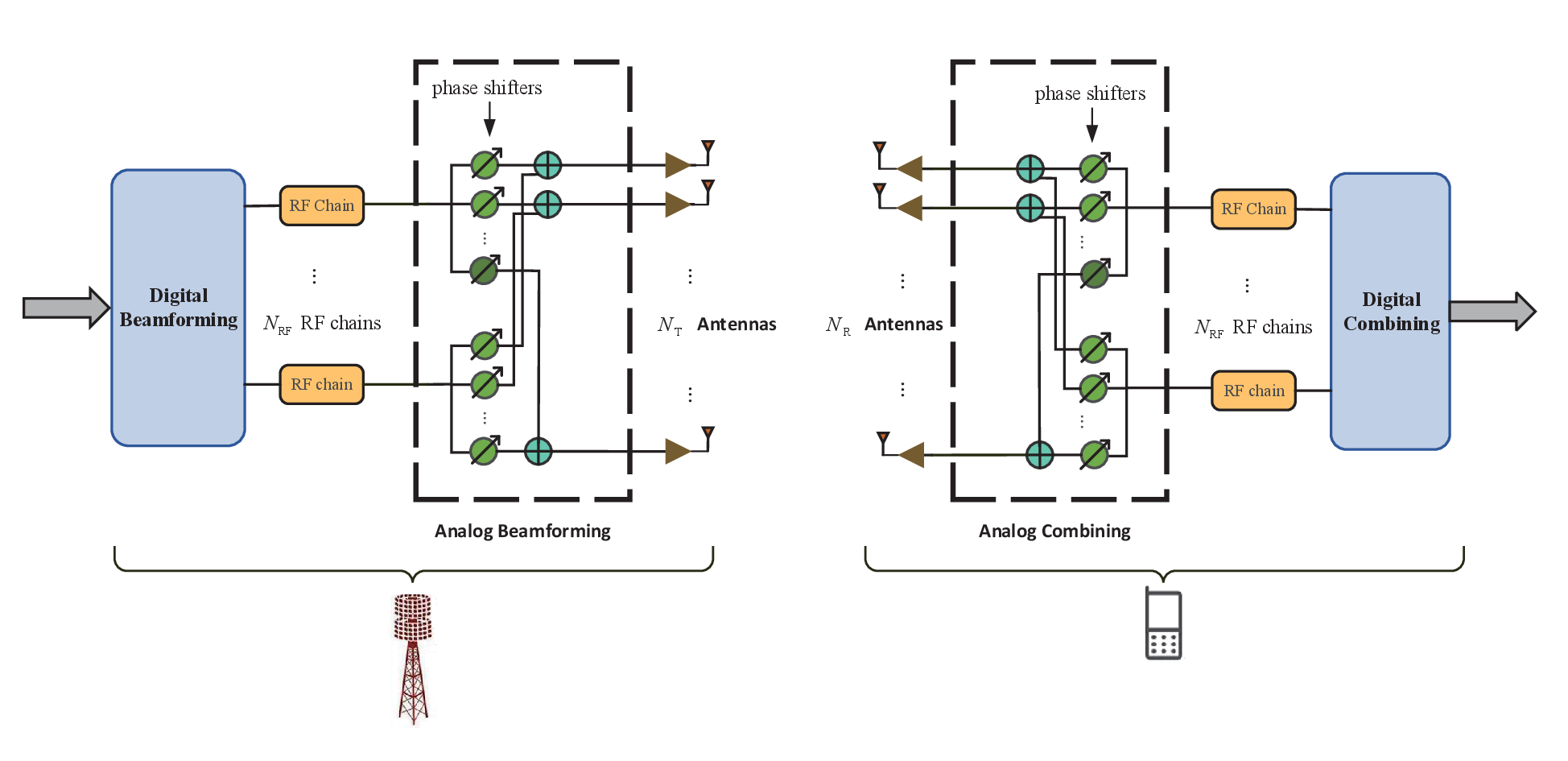}
\caption{Hybrid architecture for mmWave MIMO transceiver.}\label{SystemModel}
\end{figure*}

\section{Channel Modeling and CSI Acquisition}\label{SecChannelModeling}

In this section, we first briefly review the state-of-the-art channel models for mmWave massive MIMO systems, followed by the elaboration on the dedicated transceiver architecture. At the end of this section, an enhanced-mmWave massive MIMO system with reconfigurable intelligent surface (RIS) is briefly discussed.

\subsection{MmWave Massive MIMO Channel Modeling}

It has been shown that mmWave massive MIMO channels follow the Saleh-Valenzuela model~\cite{BOOK2007,Access2013}, which can be further divided into the narrowband model~\cite{2010AACC} and the wideband model~\cite{P_Schniter}. The narrowband model is a summation of the product of several transmitting channel steering vectors and receiving channel steering vectors, where each element involved in the summation corresponds to a channel multipath component (MPC). We consider a multiuser mmWave massive MIMO system comprising a base station (BS) and $U$ users. Let $N_{\textrm{T}}$ and $N_{\textrm{R}}$ denote the number of antennas at the BS and each user, respectively. Denote $\boldsymbol{H}_u\in{\mathbb{C}^{N_{\textrm{T}}\times{N_{\textrm{R}}}}}$ as the channel matrix between the BS and the $u$th user as
\begin{equation}
\boldsymbol{H}_{u} = \sqrt{\frac{N_{\textrm{T}}N_{\textrm{R}}}{L_u} } \sum_{i=1}^{L_u} g_{u,i} \boldsymbol{\alpha} (N_{\textrm{T}},\theta_{u,i}) \boldsymbol{\alpha}^{H} (N_{\textrm{R}},\varphi_{u,i})
\end{equation}
where $L_{u}$ and $g_{u,i}$ denote the total number of resolvable paths and the channel fading coefficient of the $i$th path for the $u$th user, respectively. The steering vector $\boldsymbol{\alpha}(N,\theta)$ is defined as
\begin{equation}
\boldsymbol{\alpha}(N,\theta)=\frac{1}{\sqrt{N}}\left[1,e^{j\pi\theta},...,e^{j\pi\theta(N-1)}\right]^{T}.
\end{equation}
Denote the angle of arrival (AoA) and angle of departure (AoD) of the $i$th path of the $u$th user as $\vartheta_{u,i}$ and $\varphi_{u,i}$, respectively, which are uniformly distributed over $[-\pi,\pi)$. Then $\theta_{u,i} \triangleq \sin{\vartheta_{u,i}}$ and $\phi_{u,i} \triangleq \sin{\varphi_{u,i}}$ if the distances between adjacent antennas at the BS and the users are with half-wave length. Since the non-line-of-sight (NLOS) MPCs are usually much weaker than the line-of-sight (LOS) MPCs, the mmWave MIMO channel is sparse in the angle domain, which only has a small number of significant entries. However, due to the channel power leakage caused by the limited resolution of phase shifters, this sparse property is not ideal~\cite{TSP2018WenyanMa}, which brings the challenge for CSI acquisition.

While the narrowband model assumes the same delay for different MPCs, the wideband model further considers the different delays of MPCs~\cite{RappaportTCOM2015}. To tackle the frequency-selective fading caused by the multipath delay spread in the wideband channel, orthogonal frequency division multiplexing (OFDM) with $K$ subcarriers is typically used to convert a wideband channel into multiple narrowband channels. Denote $\boldsymbol{H}_u^k\in{\mathbb{C}^{N_{\textrm{T}}\times{N_{\textrm{R}}}}}$ as the channel matrix on the $k$th subcarrier between the BS and the $u$th user and can be expressed as
\begin{equation}
\boldsymbol{H}_{u}^k = \sum_{d=0}^{D-1} \boldsymbol{H}_{u,d}e^{-j\frac{2\pi kd}{K}},
\end{equation}
where $D$ denotes the number of delay taps of the channel. The channel matrix at the $d$th delay tap can be expressed as
\begin{equation}\small
\boldsymbol{H}_{u,d}=\sqrt{\frac{N_{\textrm{T}}N_{\textrm{R}}}{L_u} }\sum_{i=1}^{L_u} g_{u,i} p_{u}(dT_s-\tau_{u,i}) \boldsymbol{\alpha} (N_{\textrm{T}},\theta_{u,i}) \boldsymbol{\alpha}^{H} (N_{\textrm{R}},\phi_{u,i})
\end{equation}
where $p_{u}(t)$, $T_s$, and $\tau_{u,i}$ denote the pulse shaping, the sampling
interval, and the delay of the $i$th path for the $u$th user, respectively.

As an extension of the wideband channel model, a practical model in 3GPP TR 38.901 replaces each channel MPC with a channel cluster, which further includes a number of channel rays with some minor angel offsets. The 3GPP TR 38.901 reference implementation, known as the quasi deterministic radio channel generator (QuaDRiGa)~\cite{TAP2014}, has provided a system-level software to simulate realistic radio channels including mmWave channels.

Based on the wideband model, the beam squint effect, which is caused by different delays of the received signals from different receiving antennas, has been investigated~\cite{B_WangbeamSquintTSP2018,B_Wang}. However, most research works neglect the beam squint effect by assuming that the received signal contributed by the same MPC has the same delay for different antennas. Sometimes it is necessary to further consider the detailed aspects of the mmWave MIMO channel model so that the designed mmWave system can better match the practical scenarios.

\begin{figure*}[!t]
\centering
\includegraphics[width=5.8in]{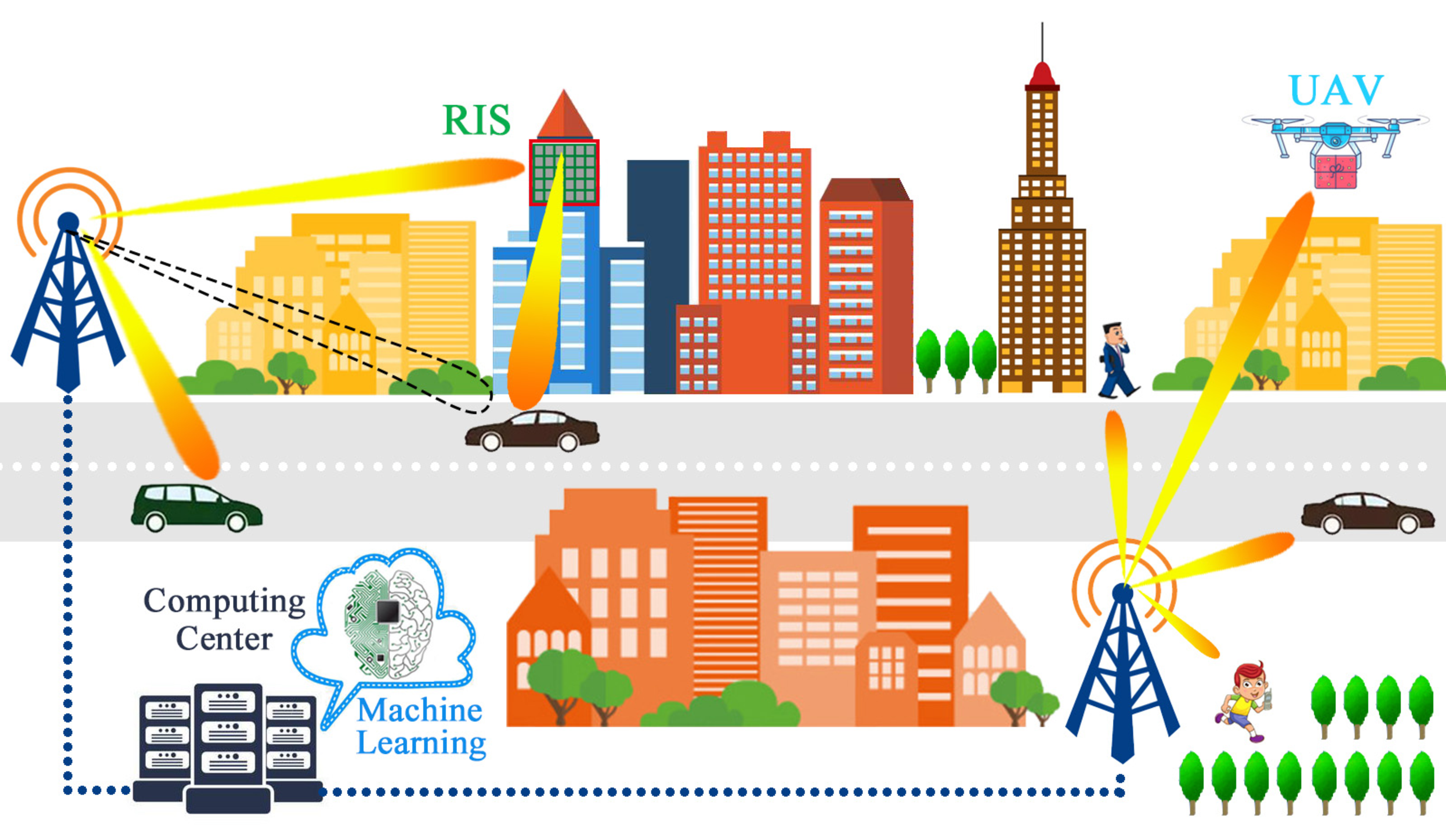}
\caption{Illustration of mmWave massive MIMO communications system.}\label{mmWaveSystem}
\end{figure*}

\subsection{MmWave MIMO Transceiver Architecture}

To balance complexity and performance, hybrid architecture, as shown in Fig.~\ref{SystemModel}, is widely used in mmWave MIMO transceiver, where several antennas share one RF chain and signal processing is partially in the digital and partially in the analog domains. Properly designed hybrid beamforming and combining, with much lower complexity, can approach the achievable rate performance of fully-digital beamforming~\cite{O_E_Ayach}. Different from the fully-digital beamforming that needs the same number of RF chains as that of the antennas, hybrid beamforming uses much fewer RF chains than the antennas and substantially reduces the hardware costs. From Fig.~\ref{SystemModel}, hybrid beamforming typically includes analog beamforming and digital beamforming, where the former generates highly directional beams based on large antenna arrays and the latter mitigates the interference among parallel data streams supported by multiple RF chains. Analog beamforming connects RF chains to antennas by the phase shifters, switches, or electromagnetic lens.

\begin{figure*}[t]
\centering
\includegraphics[width=5.2in]{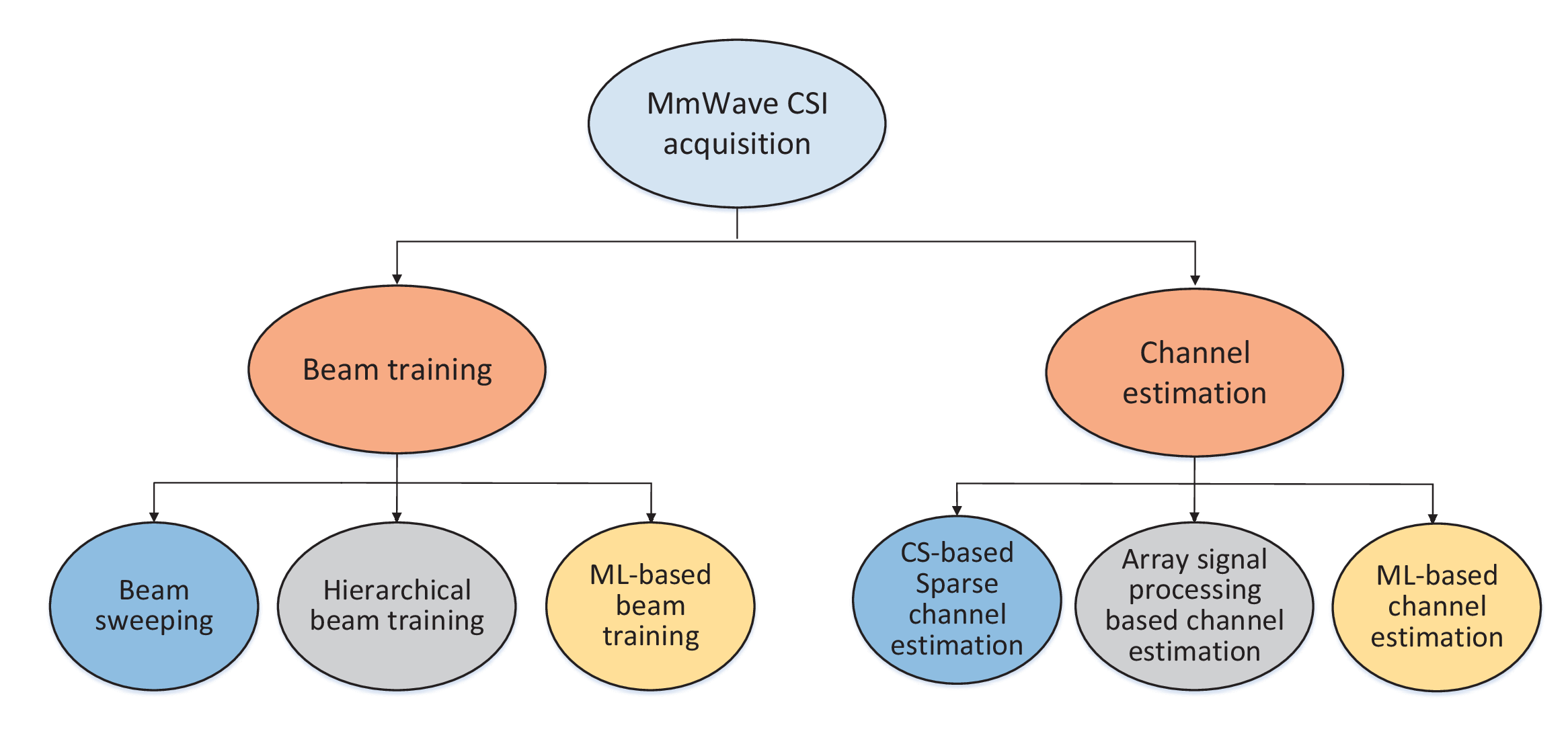}
\caption{CSI acquisition for mmWave massive MIMO.}\label{Category}
\end{figure*}

There are two different connection modes for the phase shifters: full-connection and partial-connection modes. In the full-connection mode, each RF chain connects to all antennas while in the partial-connection mode, each RF chain only connects to a subset of antennas and therefore can further reduce the hardware complexity but with some performance degradation. The phase shifters can only change the phases of signals usually with limited resolution. To further reduce the hardware complexity, low-cost switches with only on-and-off binary states, can be used to partially replace the phase shifters~\cite{Access2016,JSTSP2018HardwareEfficient}. But the time efficiency as well as the signal jitter when switching between two states should also be considered. These hardware constraints bring new challenges on CSI acquisition comparing with other communication systems. We will focus on CSI acquisition for mmWave communications in this article, as shown in Fig.~\ref{Category}.

\subsection{RIS to Enhance Signal Coverage}

Recently, RIS has been proposed to improve signal coverage~\cite{S_Hu,Access2019,MDRenzoJSAC2020,LDaiMDRenzoAccess2020}. The RIS, sometimes also known as the intelligent reflecting surface (IRS), can reflect the incident wireless signal by changing its amplitude and phase, which functions similarly as a mirror to reflect the incident light. Therefore, it can help cover the area without the LOS channel path, as illustrated in Fig.~\ref{mmWaveSystem}, which well compensates for the deficiency of the mmWave MIMO signal that mostly relies on the directional transmission. In mobile mmWave wireless communications with ground vehicles~\cite{mmWaveVehicular2016CommMag} or unmanned aerial vehicles (UAVs)~\cite{Xiao2016UAVmmWave}, the effective channel links between the users and the BS may be blocked by the buildings or other users. In this context, the RIS can effectively reflect the signal and recover the link. Therefore, the RIS can improve the reliability of the mmWave communications. But different from wireless relays, RIS is generally implemented by low-cost hardware, such as meta-material antennas, positive intrinsic-negative diodes, field effect transistors, and can passively reflect the signal in the RF frontend without any baseband signal processing capabilities. The RIS introduces additional channel links between the transmitter and the RIS as well as those between the RIS and the receiver~\cite{wang2020compressed,RIS2020SPAWCzhang}. As a result, the CSI acquisition in the RIS-assisted mmWave massive MIMO system becomes much more complicated and is different from that in mmWave systems. Since the RIS is a recent emerging technique for mmWave communications, there is no much mature work in CSI acquisition for RIS-assisted mmWave massive MIMO system and therefore we will skip the detailed discussion in this work.

\section{Beam Training}



Beam training is a process to find a pair of transmit and receive beams that best align with the strongest MPCs of the mmWave MIMO channel. Sometimes beam training is also called \textit{beam alignment}~\cite{CLiuJSAC2017}. For the mmWave massive MIMO using electromagnetic lens that generally function as a DFT transformation from the angle space to the beamspace, beam training is also called beam selection. Codebook-based beam training is a popular and general method~\cite{JSAC2009,TC2017CommonCodebook,XiaoCodebook2017TWC,K_Chen}. For this method, a codebook is first established at the transmitter and the receiver, and each codeword in the codebook, similar to the channel steering vector, generates a beam. During the beam training, we use a pair of codewords selected from the codebooks at the transmitter and receiver, respectively, to generate a pair of beams and then measure the received signal power. The pair of codewords corresponding to the largest received signal power is identified as the result of beam training. As shown in Fig.~\ref{Category}, the beam training approaches can be categorized into three branches, including beam sweeping, hierarchical beam training, and ML-based beam training. In brief, beam training identifying the best transmit and receive beams can avoid the challenge coming from estimating a high-dimension mmWave MIMO channel matrix in the scale of the number of the antennas. With beam training, we only need to estimate a low-dimension equivalent channel matrix in the scale of the number of the RF chains.

\begin{figure*}[!t]
\centering
\includegraphics[width=4.5in]{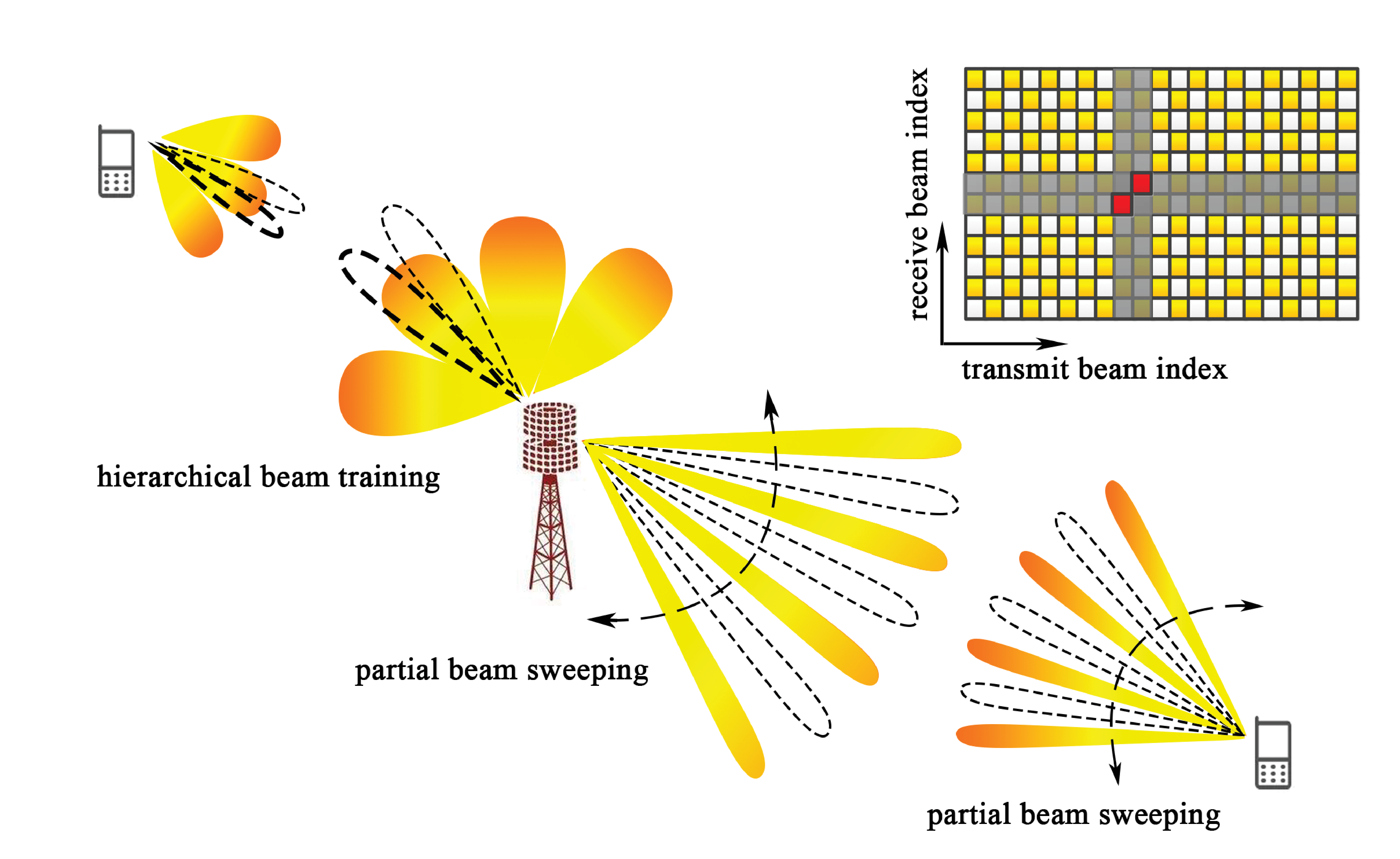}
\caption{Illustrations of partial beam sweeping and hierarchical beam training.}\label{BeamTraining}
\end{figure*}

\subsection{Beam Sweeping}

The most straightforward approach on beam training is testing all pairs of codewords to find the best one, which exhaustively selects each pair of codewords one by one and is thus very similar to sweeping different angles in the space. As a result, this approach is called beam sweeping~\cite{CST2019BeamManagement}. After beam sweeping, we can find the best pair of codewords and use them for transmit and receive beamforming. However, we may generate even-better fine-grained  beams by further performing angle estimation based on the best pair of codewords~\cite{EMHofstetterTIT1969}, since the codewords used in the beam sweeping usually point at coarse-grained discrete angles.

To reduce the overhead of beam sweeping, one scheme only uses a subset of codewords for the beam sweeping, which is called partial beam sweeping~\cite{X_Sun}. As shown in Fig.~\ref{BeamTraining}, the transmit and receive codewords can form two-dimensional grids, where each grid represents a pair of transmit and receive codewords. The partial beam sweeping divides the beam training into two stages for initial test (INTS) and additional test (ADTS). While the original beam sweeping exhaustively tests all grids in Fig.~\ref{BeamTraining}, the partial beam sweeping method only tests half or fewer of all the grids in the stage of INTS. By exploiting the coherence of different grids caused by the mmWave channel power leakage, the neighboring two columns and two rows with the largest average energy are identified, where the two tested grids on the cross of the identified two columns and two rows are predicted as the best codeword pair or best grid. Then the untested grids in the neighbor of the predicted best grid are tested in the ADTS, which only needs a very small training overhead \cite{X_Sun}. Furthermore, to break down the constraint that the grid selection in the INTS should be equally spaced, the probabilistic selection is introduced for the grid selection so that the number of the selected grids in the INTS can be set randomly. Compared with the beam sweeping, the partial beam sweeping can substantially reduce the beam training overhead with small sacrifice of SE performance.

Aside of directly reducing the training overhead, another option is to consider more efficient use of total training energy budget. In~\cite{MLiTWC2019}, the beam sweeping is divided into two stages. In the first stage, all beam pairs are tested in the same way as the beam sweeping, but a set of less favorable pairs determined from the received signal profile is then eliminated. In the second stage, an extra measurement is taken for each of the survived pairs and the best beam pair is determined by comprehensively considering these two stages.

Auxiliary beams can also be used to assist the beam sweeping as well as the beam tracking. Different from  using only one beam for beam sweeping, multiple beams with one main beam and some auxiliary beams are used together for beam sweeping. Since we can compare the power between the main beam and the auxiliary beams, estimation of channel AoA and AoD can be improved~\cite{DZhuTWC2017}. Auxiliary beams are also capable of tracking the angle variations in mobile mmWave communications scenario, where high-resolution angle tracking strategies can be designed~\cite{DZhuTWC2018}.

\subsection{Hierarchical Beam Training}

Another approach to reduce the beam training overhead is hierarchical beam training, which uses a fast codeword selection strategy based on a hierarchical codebook \cite{S_Noh,Z_Xiao,TC2-2017,GYLiJSAC2017,KangjianChen2021CL}. The hierarchical codebook has multiple layers with increasing numbers of codewords from the upper to the lower layers. That is, the upper layer consists of a small number of low-resolution codewords that generate wide beams and the lower layer has a large number of high-resolution codewords that generate narrow beams with highly directional beam gain. As shown in Fig.~\ref{BeamTraining}, the hierarchical beam training usually first tests the mmWave channel with some wide beams generated by low-resolution codewords at the upper layer and then narrows down the beam width layer by layer until a best codeword at the bottom layer is obtained.

Many works focus on the codeword design under different hardware constraints, including the limited resolution, constant envelop of phase shifters, and the limited number of RF chains. To design a codeword with a wide beam coverage, the simplest scheme is to power off some antennas; but will consequently reduce the total radiation power and affect the signal coverage. A better scheme is to divide a large antenna array into several subarrays \cite{Z_Xiao}, where each subarray generates a beam and we optimize the weighted summation of the subarray beams to approximate the objective beam pattern. To further improve the performance of beam pattern for the generated beams under the hardware constraints, e.g., reducing the mainlobe fluctuations or enlarging the sidelobe attenuation of the generated beam, we may design ideal codewords in the first step by ignoring the hardware constraints and then generate practical codewords considering the hardware constraints to approximate the ideal codeword by alternative minimization in the second step \cite{K_Chen}. For the non-convex codeword design problem under the constraints of the transition band and the ripple on mainlobe or sidelobes, an efficient algorithm is proposed and closed-form expressions of the beam training error rate is derived~\cite{TC2-2017}. To support simultaneous beam training for multiple users and therefore improve the efficiency of multiuser beam training, adaptive hierarchical codebook based on multi-mainlobe codewords can be further considered, where each mainlobe can be pointed at a spatial region that a user may be probably located in~\cite{HierarchicalBeamTraining2019Globecom,Chenhao2020TWCcodebook}. As such, the total training overhead can be substantially reduced.

\subsection{ML-Based Approaches}
Recently, ML has been introduced for the mmWave beam training by exploiting the temporal, frequency or spatial correlation of the mmWave communications. Two main branches of ML for beam training includes supervised learning and reinforcement learning (RL), where the labeled training dataset is required by the former but not by the latter. The RL enables an agent to learn how to take a good action on the environment aiming at getting the largest rewards from the environment, by continuous exploration and exploitation.

\subsubsection{Supervised Learning-based Approaches}
The commonly used methods for supervised learning includes the methods based on neural network (NN), such as deep NN (DNN), convolutional NN (CNN), recurrent NN (RNN). To reduce the overhead of beam sweeping, a beam alignment method only testing partial beams can be used to predict the beam distribution and the best beam pair based on the NN, where the NN is trained offline to exploit the channel spatial correlation using simulated mmWave channel data~\cite{WenyanMaWCL2020}. To improve the efficiency of beam training, beams with different widths, e.g., wide beams and narrow beams, can be used. Given the wide beam measurements, a super-resolution-inspired deep learning method based on the CNN can be used to estimate the beam qualities of narrow beams and therefore select the best narrow beam for data transmission. By capturing the spatial correlation as well as the temporal correlation, a beam qualities prediction model based on a convolutional long short term memory (LSTM) network is used to estimate the current beam qualities from the historical data~\cite{HEchigoTVT2021}. In fact,  we may treat a pair of transmit and receive beams as a pixel of an image, where the beam training can be treated as an image reconstruction problem. Based on the off-line eigen-beam extraction from the original beam-domain receive power map (BDRPM) and partial beam measurements, we can reconstruct a full BDRPM by the DNN~\cite{CLinVTC2019}.

To assist the beam training, additional information including the positions and orientation of users~\cite{KSatyTVT2019,LocationOrientationICC2020}, situational awareness~\cite{Y_Wang}, and those coming from sub-6GHz channel~\cite{BlockagePrediction2020TCOM} and light detection and ranging (LIDAR) sensors~\cite{LIDAR2019WCL} can be employed. In~\cite{KSatyTVT2019}, a pair of transmit and receive beams can also be considered as a fingerprint at a particular position. Based on a NN trained by a fingerprint database, we can predict the candidate beam pairs given the position of the users, which will only incur small beam training overhead. In~\cite{LocationOrientationICC2020}, the position and orientation are input to a DNN that essentially functions as a multi-labeled classifier to obtain the probability of each beam pair being the best one under indoor scenarios. By collecting and exploiting the situational awareness information, e.g., the positions and types of the receiver and its neighbors, the mmWave beam training can be formulated as a classification problem, where the ensemble learning methods can be leveraged to design efficient and robust schemes~\cite{Y_Wang}. It is well known that the sub-6 GHz wireless channels have better propagation condition than the mmWave ones. If sub-6 GHz channels are available, the mapping functions to predict the best mmWave beam pair from the sub-6 GHz channels can be better modeled by the DNN than the analytical methods~\cite{BlockagePrediction2020TCOM}. In autonomous driving, LIDAR is a sensor widely equipped for high-resolution mapping and position, where the LIDAR sensor data can be used to train the CNN to find a set of beams for mmWave communications~\cite{LIDAR2019WCL}.

\subsubsection{RL-based Approaches}
Different from the supervised learning that relies on the labeled training dataset, the RL does not need such kind of information and therefore is more flexible. Aiming at obtaining the largest rewards from the environment, an agent learns how to take a good action towards the environment through continuous exploration and exploitation. As a straightforward approach of RL, multi-armed bandit (MAB) can be applied for mmWave beam training. MAB addresses a sequential decision problem that an agent selects an arm with a trade-off between exploitation and exploration to maximize the expected reward. Each pair of transmit and receive beam can be regarded as an arm of the MAB model. Then the beam training can be formulated as the problem of finding a best arm, which can be typically solved by upper confidence bound (UCB) algorithms. In~\cite{M_Hashemi}, the correlation and unimodality properties across beam training are exploited, which inspires the proposed unimodal beam alignment algorithm with the asymptotically optimal performance. In~\cite{VVaAccess2019}, the beam training is divided into two stages, including the beam pair selection stage and the beam refinement stage. In the first stage, coarse beam directions is identified by the proposed greedy UCB and risk-aware greedy UCB algorithms. In the second stage, a modified optimistic optimization algorithm further refines the results from the first stage. Considering the multipath propagation of mmWave channels, $M$ arms with $M$ largest UCBs instead of only one arm are selected at each time slot, where the transmission rate is iteratively updated according to the feedback~\cite{MChengICC2019}.

In addition to the UCB algorithm, the Bayesian methods can also be used to solve the beam alignment problem based on the MAB model. The linear Thompson sampling can be integrated with sparse Bayesian learning and Kalman filters for smart exploration of feasible actions~\cite{MBBoothCL2019}. In~\cite{MHussainGLOBECOM2019}, a second-best preference policy is proposed to restrict the value function and select the beam pair with the current sub-optimal priority to achieve the best balance between exploitation and exploration.

Besides the UCB and Bayesian algorithms, a fast machine learning method based on contextual information is proposed to use rough user location and aggregate the received data to learn and continuously adapt to the actual channel environment~\cite{GHSimACMTN2018}, while a hierarchical beam alignment method exploiting the inherent correlation among beams determines the optimal beam with accelerated BA process and reduced beam training overhead~\cite{WWuTWC2019}.

Compared to the MAB model, it is more sophisticated to model the beam training as a Markov decision process, where Q-learning methods are typically employed. In~\cite{SKimICTC2020}, the estimation of beam steering angle is divided into two phases. In the first phase, Q-learning based beam tracking finds the optimal beam. In the second phase, several beams neighbouring the optimal beam found in the first phase are tested, where the tested beam pair with the largest received signal strength is then selected for the modified auxiliary-beam-pair-based angle estimation. In fact, the Q table in the Q-learning methods can be well approximated by the DNN, which is named as deep Q-network (DQN). In~\cite{JZhangGLOBECOM2019}, the DQN is used to learn the change of the mmWave communication channels and intelligently train the beams with a low overhead.




\section{Channel Estimation}

Channel estimation, as the other category of CSI acquisition, aims at accurately estimating the mmWave massive MIMO channels. As shown in Fig.~\ref{Category}, the mainstream channel estimation methods can be divided into three branches, including CS-based sparse channel estimation, array signal processing based channel estimation, and ML-based channel estimation.

\subsection{CS-based Sparse Channel Estimation}

Extensive works on channel measurement and modeling have demonstrated that the scattering environment in mmWave frequencies is sparser than its lower frequency counterpart, such as in sub-6 GHz channel. Therefore, we can exploit the channel sparsity by formulating channel estimation as a sparse recovery problem and solve it using the CS algorithms. In general, we first obtain the AoDs/AoAs and then estimate the channel gains. Denote $\mathbf{z}$ as a sparse vector, where its locations and values of the non-zero entries indicate the quantized AoDs and AoAs and path gains, respectively. The performance loss caused by the quantization error proves to be minor with a fine-grain enough candidate set for AoDs and AoAs~\cite{A_Alkhateeb}. Due to the channel sparsity, the number of non-zero entries in $\mathbf{z}$ is much smaller than its dimension.

The orthogonal matching pursuit (OMP) algorithm is popular for sparse channel estimation for mmWave wireless system~\cite{A_Alkhateeb,alkhateeb2015compressed,lee2016channel,wang2020compressed}. During the downlink channel training, random beamforming and projection can be used to estimate the mmWave channel, where the OMP algorithm is adopted to estimate the quantized AoDs and AoAs corresponding to the largest entries of $\mathbf{z}$ and a lower bound on the achievable rate as a function of the CS measurements is provided~\cite{alkhateeb2015compressed}. However, the proposed method in \cite{alkhateeb2015compressed} only estimates the LOS channel path of the mmWave channel, which does not make full use of mmWave channel resources. To solve this problem, OMP-based estimation methods for multiple channel paths are proposed~\cite{A_Alkhateeb,lee2016channel}. Different from the random beamforming adopted in \cite{alkhateeb2015compressed}, a hierarchical codebook with adaptive CS algorithm to recover $\mathbf{z}$ is designed in \cite{A_Alkhateeb}, which can invoke hybrid processing during the pilot transmission and thus yield more efficient beam patterns to capture the non-zero entries in $\mathbf{z}$. The AoD, AoA, and the gain of each channel path are iteratively distilled from the residual signal, where the residual signal is iteratively updated by subtracting the components contributed by the previously estimated paths. In~\cite{lee2016channel}, the OMP-based channel estimation is evaluated with a redundant dictionary, which consists of array response vectors with finely quantized angle grids non-uniformly distributed in $[0, \pi]$. The lower and upper bounds of the sum-of-squared errors of the proposed method are analytically derived, while the pilot vectors are designed to minimize the coherence of the sensing matrix. Note that the receiving noise in mmWave system is not white since it is multiplied by the hybrid combiner at the receiver. In~\cite{javier2018frequency}, a simultaneous-weighted-OMP algorithm exploits the common support among different subcarriers, where the spatial noise components are simultaneously whitened by Cholesky factorization of the hybrid combiner and the most likely support index in the sparse beamspace channels can be estimated more accurately before whitening. Since the OMP algorithm only saves the best candidate atom having the largest projection with the residue at each iteration, it estimates the dominant channel entries sequentially and greedily, which cannot guarantee the globally optimal performance. In~\cite{tao2019regularlized}, a regularized multipath matching pursuit algorithm better than OMP is proposed for sparse channel estimation, where multiple candidate atoms instead of only one atom is kept at each iteration and a regularization step is introduced to reduce the computational complexity.

Sparse channel estimation for mmWave massive MIMO with electromagnetic lens structure draws the research interest, since the electromagnetic lens with switches have lower hardware complexity than the phase shifter networks. In~\cite{dai2017tras}, mmWave system with 2D lens is considered, where the structural characteristics of beamspace channel show that the support of the sparse beamspace channel is uniquely determined by its largest entry. Therefore, only one projection is needed to determine the beamspace channel support of each path, which can substantially reduce the computational complexity. Better performance can be achieved by the proposed method in~\cite{dai2017tras}, especially in the low signal-to-noise ratio (SNR) region, compared to the classical CS algorithms. The mmWave system with 3D lens is considered in~\cite{gao2016estimation,ma2017channel}. It is shown that the most power of the beamspace channel matrix is concentrated on a much smaller submatrix, which inspires the adaptive support detection method to model the beamspace channel support as a rectangle, with the adaptively modified width and length of the rectangle\cite{gao2016estimation}. It is further demonstrated in~\cite{ma2017channel} that the main power of the beamspace channel is more concentrated on an area of dual crossing than the rectangle, which brings forward a dual crossing channel estimation method to iteratively refine the selection of dominant entries until the stop condition is met. At each iteration, two adjacent columns and rows with the largest channel power are identified to form the area of dual crossing, where the selection of the dominant entries is adjusted by deleting the weakest outer entry at dual crossing corners and then adding a new entry outside of the strongest outer entry.

To mitigate the performance loss caused by the quantization error, we may increase the resolution of the AoD/AoA candidate set, which however incurs high hardware complexity. To tackle this problem, off-grid super-resolution channel estimation based on gradient descent~\cite{C_Hu}, atomic norm minimization~\cite{YingmingTsaiTCOM2018}, Dirichlet kernel~\cite{anji2019channle}, and distributed grid matching pursuit~\cite{dai2016estimation} has been investigated. Inheriting the formulated sparse recovery problem, the intractable $l_{0}$-norm minimization is equivalently transformed into the minimization of a surrogate function~\cite{C_Hu}. By resorting to the iterative reweighting method, the above equivalent problem is then solved with the AoDs and AoAs updated based on gradient descent. If the initialized channel sparse level is set larger than the genuine one, we can find the genuine number of paths by iteratively pruning those false paths. To reduce the computational complexity, singular value decomposition based preconditioning can be further used by properly initializing the AoDs and AoAs. Channel estimation can also be formulated as an atomic norm minimization problem, which avoids discretizing the angle spaces of the AoA/AoD into grids and therefore achieves high-accuracy channel estimation~\cite{YingmingTsaiTCOM2018}. Further reformulating the atomic norm minimization problem as a semi-definite program, alternating direction method of multipliers can be employed where each iteration only needs closed-form computation. By assuming that the beamspace parameters can be any value in beamspace domain and do not restrict to predefined discrete angles, implicit Dirichlet kernel structure in the Fourier domain is used, where the OMP algorithm is employed to find the maximum of the Dirichlet kernel peaks instead of recovering all the non-zero entries in the virtual DFT domain~\cite{anji2019channle}. In~\cite{dai2016estimation}, a channel estimation method based on distributed grid matching pursuit is developed to iteratively detect and adjust the channel support, where the adaptive measurement matrix is determined by the most possible channel paths in the outer loop iteration and the AoAs and AoDs estimation associated with the LOS path is improved with the grid matching strategy in the inner loop iteration.

\subsection{Array Signal Processing Based Channel Estimation}

\begin{figure*}[t]
\centering
\includegraphics[width=7.2in]{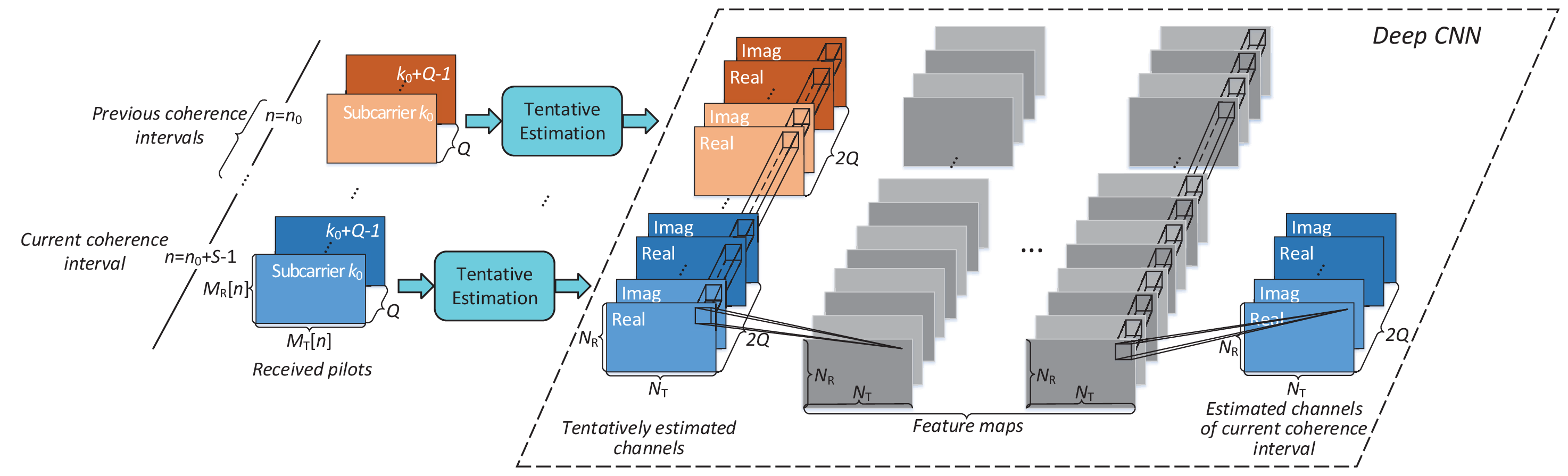}
\caption{Framework on deep CNN-based channel estimation.}\label{DeepCNN}
\end{figure*}

Array signal processing technique, originally proposed for the AoA estimation in radar system with a large antenna array, can also be applied for mmWave massive MIMO channel estimation. The estimating signal parameters via rotational invariance techniques (ESPRIT) algorithm and multiple signal classification (MUSIC) algorithm are the classical AoA estimation algorithms used in radar systems. But different from the radar system, the mmWave massive MIMO typically employs a hybrid beamforming architecture where the RF chains are much fewer than the antennas, meaning that we can only obtain a low-dimension signal from the RF chains instead of directly getting a high-dimension signal from the frontend antennas. Compared to the high-dimension mmWave MIMO channel matrix to be estimated, the low-dimension signal from the RF chains is not enough.

Channel estimation methods based on the ESPRIT algorithm are proposed in \cite{liao20172d,zhang2017channel,zhang2017channel2,W_Ma_a}. The training
signals at both the BS and each user are designed to obtain the low-dimensional effective channel~\cite{liao20172d}, while the spatial smoothing and the forward backward averaging techniques are adopted to alleviate the impact of coherent signals caused by multiple AoAs or AoDs close to each other. Moreover, the channel matrix is converted to a real matrix so that the computational complexity can be reduced. To avoid the information loss from the high-dimension signal to the low-dimension signal, a straightforward method is to power off some antennas so that the time slots for channel estimation equals that of the antennas, by which we estimate the AoAs and AoDs through a submatrix of the high-dimension mmWave MIMO channel matrix and eventually combine the submatrices together to obtain an estimate of the full channel matrix. Two methods based on the beamspace ESPRIT algorithm are proposed in~\cite{zhang2017channel,zhang2017channel2}, where the hybrid precoding matrix and combining matrix are designed as the DFT matrix so that the beamspace ESPRIT algorithm can be directly applied. However, the hybrid precoding matrix and combining matrix can only be partial DFT matrix, since the time slots for channel estimation are less than the total antennas. Consequently, the AoAs and AoDs to be estimated are restricted in a small range, indicating that the aforementioned methods are not generally applicable for the estimation of any AoAs and AoDs. Note that powering off antennas will reduce the total radiation power and therefore reduce the signal coverage. A better method is to turn off less antennas, e.g., only one antenna~\cite{W_Ma_a}. To achieve high-resolution estimation of AoAs and AoDs using the ESPRIT method, multiple stages of pilot transmission can be considered. AoAs and AoDs can be first estimated by a two-dimensional ESPRIT scheme and then paired, based on which the channel gain can be estimated by the least squares (LS) method. To further reduce the pilot overhead, a one-dimensional ESPRIT and minimum searching-based scheme only needing two stages is proposed, where the AoAs are first estimated and then the AoDs are obtained by searching the minimum value within the identified mainlobe. Besides, hybrid beamforming and combining matrices for the pilot transmission can be designed to yield almost equal received powers for any AoA and AoD so that robust super-resolution channel estimation can be achieved.

Channel estimation methods based on the MUSIC algorithm are proposed in~\cite{guo2017millimeter,wu2018long}. The beamspace 2D MUSIC algorithm in~\cite{guo2017millimeter} estimates the AoAs and AoDs of the incoherent signals and then the LS method is adopted to estimate the channel gain for each path. Similar to~\cite{zhang2017channel}, the hybrid precoding and combining matrices are partial DFT matrices, indicating that the beamspace 2D MUSIC algorithm can only estimate AoAs and AoDs in a small range. For the coherent signals, the beamspace 2D MUSIC algorithm is also used~\cite{wu2018long}. Since the signals are coherent, different signals can merge into one signal, which decreases the number of independent sources. In this context, the received signals are multiplied by an exchange matrix so that the number of antennas in subarrays equals the total number of array entries. However, the hybrid precoder and combiner can only be partial DFT matrices, which also suffers from the drawbacks of small estimation range.

\subsection{ML-Based Solutions}
The fast development of ML motivates its application in channel estimation for mmWave massive MIMO. In particular, deep learning (DL) based methods have attracted many interests since they can extract features via multiple neural layers and accelerate the computation dramatically resorting to parallel computing. The deep neural network (DNN) can be trained to estimate the channel parameters including AoAs, AoDs, and channel gains.

ML-based channel estimation using the denoising methods are proposed in \cite{he2018deep,wei2019amp,zhang2020deep,wei2020deep,jin2019channel}. In~\cite{he2018deep}, the beamspace mmWave massive MIMO channel matrix is treated as a 2D image, where a learned denoising-based approximate message passing (LDAMP) NN integrating denoising CNN estimates the mmWave channel by subtracting the estimated residual noise from the noisy channel. To further improve the estimation performance of~\cite{he2018deep}, the denoising process is divided into several stages, mainly including the preprocessing stage and the fine processing stage~\cite{wei2019amp,zhang2020deep}. In~\cite{wei2019amp}, a learned approximate message passing (LAMP)-based network with deep residual learning for channel estimation includes two stages. In the first stage, the LAMP network exploits the beamspace channel sparsity to get a preliminary channel estimation, while in the second stage the deep residual learning further refines the coarse estimation obtained in the first stage and reduces the impact of channel noise. In~\cite{zhang2020deep}, a fully convolutional denoising approximate message passing algorithm has three stages. In the first stage the LAMP network processes the received signal and the mmWave channels; in the second stage, a fully convolutional denoising network extracts noise characteristics and obtains the noise level; and in the final stage, the feature channel is increased or decreased, respectively, by down-sampling or up-sampling. To accurately estimate the beamspace channel, the original soft-threshold shrinkage function in the LAMP network can be replaced by a derived Gaussian mixture shrinkage function, based on the prior information that the beamspace channel elements can be modeled by the Gaussian mixture distribution~\cite{wei2020deep}. Note that the above denoising-based channel estimation methods are tailored to specific noise levels. To improve the flexibility and efficiency of denoising-based channel estimation, a fast and flexible denoising CNN is proposed, where the sparse mmWave channel matrix is also treated as a natural image~\cite{jin2019channel}. By using a flexible noise level map as the input, the proposed denoising network can be used for a wide range of SNR levels. Moreover, to reduce the training and testing latency, each image in the proposed denoising network is divided into several sub-images.

In fact, the ML can also be directly used to predict the mmWave channel besides functioning as denoisers~\cite{moon2020deep,W_Ma_b,P_Dong}. In~\cite{moon2020deep}, the DNN learns the mapping function between the received omni-beam patterns and the mmWave channel, while the LSTM is adopted to track the channel. However, in practical mmWave wireless system equipped with a large antenna array, it is difficult to obtain the omni-directional signal, which requires powering off all the antennas except one antenna and consequently reduces the radiation power and signal coverage. Since the mmWave channel exhibits sparse property in beamspace, the beamspace channel amplitude can be introduced for the training of the DNN~\cite{W_Ma_b}. Then based on it, the significant entries in the beamspace channel vector can be predicted. This method is more flexible than directly predicting the positions of nonzero entries, especially when the beamspace channel is not ideally sparse. The non-zero entries of the sparse channel can be obtained simultaneously by a well trained DNN with significantly reduced running time, instead of the sequential greedy search by heuristic sparse recovery algorithms, such as OMP. Moreover, the DNN can be trained under noisy channel environment with interference, making it robust to various channel conditions. To provide accurate channel estimation close to the performance of the ideal minimum mean-squared error estimator, a robust framework on DL-based channel estimation has been proposed for mmWave massive MIMO-OFDM systems \cite{P_Dong}. The main idea is using the deep CNN to capture the spatial, frequency, and temporal channel correlation simultaneously, which is difficult for the traditional methods. Fig.~\ref{DeepCNN} illustrates the framework of deep CNN-based channel estimation, where $N_{\textrm{T}}$ and $N_{\textrm{R}}$ are the numbers of antennas at the transmitter and the receiver, respectively, and $M_{\textrm{T}}[n]$ and $M_{\textrm{R}}[n]$ are the numbers of beamforming and combining vectors, respectively, for pilot transmission in the $n$th coherence interval. As shown in Fig.~\ref{DeepCNN}, the received pilots at $Q$ subcarriers in the current coherence interval are first processed by the tentative estimation module. Then the tentatively estimated channels together with those estimated in the previous $S-1$ coherence intervals are input into the developed deep CNN to obtain the estimated channels at $Q$ subcarriers of the current coherence interval. Through proper design, the CNN is able to capture the channel temporal correlation efficiently with low complexity even without incorporating the long short-term memory architecture. Besides improving the estimation accuracy and saving the running time, the proposed framework also reduces the pilot overhead. Specifically, several successive coherence intervals are grouped as a channel estimation unit, within which the pilot overhead can be reduced gradually at the cost of limited performance loss. The developed CNN-based framework is robust to different propagation scenarios even unseen in the offline training stage and without any prior knowledge of the channel conditions.

\section{Performance Evaluation}

In this section, we compare several typical approaches introduced in the previous sections in terms of SE, computational complexity, and overhead.

We consider a multiuser mmWave massive MIMO system, where a BS equipped with $N_{\textrm{T}}=64$ antennas and $N_{\textrm{RF}}=4$ RF chains serves $U=3$ single-antenna users. The mmWave massive MIMO channels are generated based on the Saleh-Valenzuela model. In Table~\ref{table_1}, we compare three beam training schemes and three channel estimation schemes, including partial beam sweeping \cite{X_Sun}, hierarchical beam training~\cite{Z_Xiao}, the MAB-based scheme~\cite{M_Hashemi}, the adaptive CS-based scheme~\cite{A_Alkhateeb}, the ESPRIT-based scheme~\cite{W_Ma_a}, and the deep CNN-based scheme~\cite{P_Dong}.

\begin{table*}[!t]
\centering
\caption{Comparisons of Different Channel Estimation or Beam Training Schemes}
\label{table_1}
\begin{tabular}{p{1.5cm}|p{3cm}|p{5cm}|p{3cm}}
\hline
\hline
~ & {\makecell{Scheme Name}} &  {\makecell{Computational Complexity}} & {\makecell{Overhead}}\\
\hline
\multirow{3}{*}{\makecell{Beam \\Training}} & \makecell{Partial beam \\sweeping \cite{X_Sun}} & $\mathcal{O}\left(U\left(\frac{N_{\textrm{T}}}{2N_{\textrm{RF}}}+2\right)+4U^3\right)$ & $2\left(\frac{N_{\textrm{T}}}{2N_{\textrm{RF}}}+3\right)$\\
\cline{2-4}
~ & \makecell{Hierarchical beam \\training \cite{Z_Xiao}} & $\mathcal{O}\left(U\log_2 N_{\textrm{T}}+4U^3\right)$ & $2U\log_2 N_{\textrm{T}}+U$\\
\cline{2-4}
~ & \makecell{MAB-based \\scheme \cite{M_Hashemi}} & $\mathcal{O}\left(N_{\textrm{T}}^3\right)$ & \makecell{Depend on \\convergence speed}\\
\hline
\multirow{3}{*}{\makecell{Channel \\Estimation}} & \makecell{Adaptive CS-based \\scheme \cite{A_Alkhateeb}} & $\mathcal{O}\left(U\log_2 N_{\textrm{T}}+4U^3\right)$ & $2U\log_2 N_{\textrm{T}}+U$\\
\cline{2-4}
~ & \makecell{ESPRIT-based \\scheme \cite{W_Ma_a}}& $\mathcal{O}\left(8UT^3+2UL^2(10L+5T)\right)$ & $T$\\
\cline{2-4}
~ & \makecell{Deep CNN-based \\scheme \cite{P_Dong}} & $\mathcal{O}\left(UN_{\textrm{T}}^2+UN_{\textrm{T}}\sum_{l=1}^{L_{c}}F_l^2N_{l-1}N_{l}\right)$ & $T$\\
\hline
\hline
\end{tabular}
\vspace{-0.3cm}
\end{table*}

\subsection{SE Performance}

\begin{figure}[t]
\centering
\includegraphics[width=3.5in]{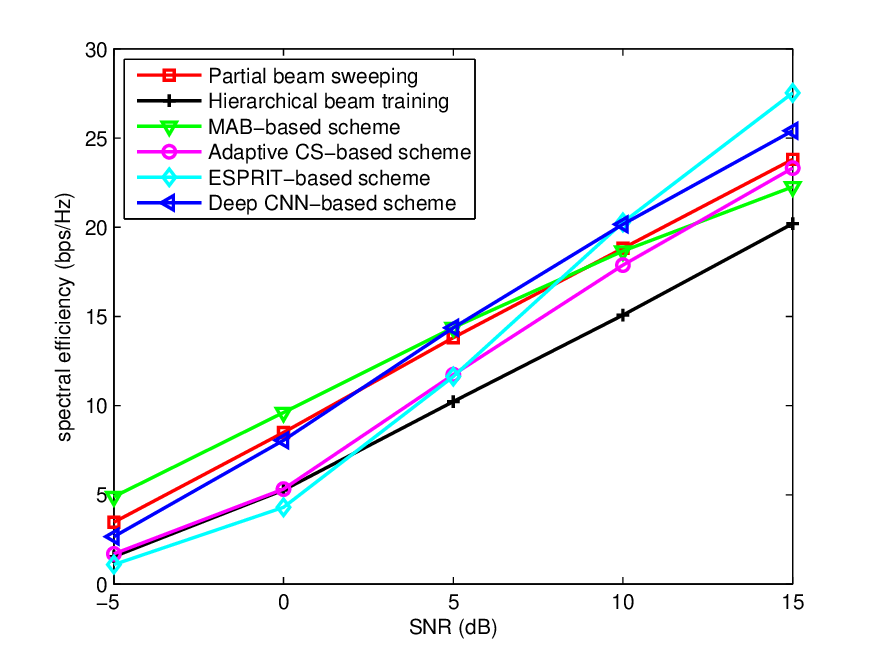}
\caption{Comparisons of spectral efficiency for different schemes in terms of SNR.}\label{SE}
\end{figure}

Fig.~\ref{SE} compares the SE at different SNRs for the six schemes. From the figure, at the low SNR region, partial beam sweeping, the MAB-based scheme, and the deep CNN-based scheme outperform the other three schemes. At $\textrm{SNR}=-5$dB, the deep CNN-based scheme has 56.8\%, 143.1\% and 73.2\% performance improvement over the adaptive CS-based scheme, the ESPRIT-based scheme and the hierarchical search-based scheme, respectively, while the partial beam sweeping-based scheme has 104.7\%, 217.4\% and 126.1\% performance improvements over the adaptive CS-based scheme, the ESPRIT-based scheme and the hierarchical beam training scheme, respectively. Since the narrow beams with large beam gain are used by both the partial beam sweeping and the MAB-based scheme, better anti-noise performance can be achieved. On the contrary, for the hierarchical beam training and adaptive CS-based schemes relying on hierarchical codebooks, wide beams with relatively small beam gain are used at the upper layer, which leads to worse anti-noise performance. The deep CNN-based scheme can train the neural network under predefined noisy condition and can guarantee better anti-noise performance than the other two schemes. At $\textrm{SNR}=15$dB, the ESPRIT-based scheme has 8.3\%, 15.7\%, 18.0\%, 23.6\%, and 36.3\% performance improvement over the deep CNN-based scheme, the partial beam sweeping scheme, the adaptive CS-based scheme, the MAB-based scheme, and the hierarchical beam training scheme, respectively. Since the ESPRIT-based scheme employs singular value decomposition to achieve super-resolution estimation of the AoA and AoD, it performs the best in the high SNR region and the worst in the low SNR region among all the schemes.

\subsection{Computational Complexity}

The computational complexity for the six schemes are compared in Table~\ref{table_1}. To compare computational complexity for different schemes, we set the parameters of the deep CNN-based scheme as \cite{P_Dong}, where $Q=1$, the number of convolution layers $L_c=10$, the side length of the filters used by the $l$th layer $F_l=3$, the number of input feature maps of the $l$th layer $N_{l-1}=2$, for $l=1$ and $N_{l-1}=64$, for $l=2,\ldots,10$, and the number of output feature maps of the $l$th layer $N_{l}=64$, for $l=1,\ldots,9$ and $N_{l}=2$, for $l=10$. By setting the number of channel paths and the length of pilot symbols to be $L=2$ and $T=8$, respectively, we can figure out that the computational complexities of partial beam sweeping, hierarchical beam training, MAB-based scheme, adaptive CS-based scheme, ESPRIT-based scheme, and deep CNN-based scheme, are in the orders of the magnitudes of $10^2$, $10^2$, $10^5$, $10^2$, $10^4$ and $10^7$, respectively. In brief, the deep CNN-based scheme has the largest computational complexity among all the schemes but can be accelerated dramatically via parallel computing.

\subsection{Overhead Comparison}

To compare the overhead for different schemes in the unit of time slots, we assume either a test of a beam pair or transmission of a pilot symbol occupies one time slot in Table~\ref{table_1}. Note that both the ESPRIT-based scheme and deep CNN-based scheme need $T$ time slots, which is independent of the numbers of the antennas, served users, and RF chains. Since both the adaptive CS-based scheme and the hierarchical beam training scheme are based on hierarchical codebooks, they have the same overhead, which is larger than that of partial beam sweeping for a small $N_{\textrm{T}}$, and vice versa.

\section{Conclusions and Open Issues}\label{SecConclusions}

In this article, we have provided an overview on CSI acquisition for mmWave massive MIMO, including beam training and channel estimation, with focus on the traditional and the machine learning based approaches. The in-depth review along with a comprehensive performance comparison demonstrates the promising prospect of mmWave massive MIMO communications from theory to practice. However, there are still many open issues for future research. Here are some of them.

\vspace{-0.4cm}

\subsection{CSI Acquisition for Mobile MmWave Massive MIMO}

Since the mmWave massive MIMO uses highly directional beams to compensate the path loss, the CSI acquisition and tracking are critical for reliable communications, especially in the mobile scenario, e.g., communications with ground vehicles or UAVs. How to efficiently manage the beams and design the channel tracking algorithms in case of channel blockage is an open issue.

\vspace{-0.4cm}
\subsection{CSI Acquisition for RIS-Assisted MmWave Massive MIMO}

The RIS introduced in Section~\ref{SecChannelModeling} can improve the signal coverage of mmWave massive MIMO by creating additional channel links. The beam training and channel estimation for RIS-assisted mmWave massive MIMO worth further investigation, especially in the dynamic wireless environment with multiple RISs. The low-cost and passive characteristic of RIS and hardware constraints should be taken into account.

\vspace{-0.4cm}
\subsection{CSI Acquisition Based on RL}

RL can interactively learn how to take a good action based on the reward and does not rely on the labeled training dataset, which well matches with the mmWave channel with short coherence time. How to effectively integrating RL in the CSI acquisition, i.e., conceiving smart beam training strategies or intelligent channel estimation schemes, deserves more studies.

\subsection{Advanced CSI Acquisition Techniques}
Although the beam training and channel estimation are presented in two separate categories in this article, they are not independent to each other. Sometimes we may combine them together, e.g., first using beam training benefit from directional transmission gain and then using advanced estimation methods coming from channel estimation category for flexible parameter estimation. More efficient and intelligent CSI acquisition techniques are interesting and worth future study.



\ifCLASSOPTIONcaptionsoff
  \newpage
\fi

\bibliographystyle{IEEEtran}
\bibliography{IEEEabrv,IEEEexample}

\begin{thebibliography}{10}
\providecommand{\url}[1]{#1}
\csname url@samestyle\endcsname
\providecommand{\newblock}{\relax}
\providecommand{\bibinfo}[2]{#2}
\providecommand{\BIBentrySTDinterwordspacing}{\spaceskip=0pt\relax}
\providecommand{\BIBentryALTinterwordstretchfactor}{4}
\providecommand{\BIBentryALTinterwordspacing}{\spaceskip=\fontdimen2\font plus
\BIBentryALTinterwordstretchfactor\fontdimen3\font minus
  \fontdimen4\font\relax}
\providecommand{\BIBforeignlanguage}[2]{{%
\expandafter\ifx\csname l@#1\endcsname\relax
\typeout{** WARNING: IEEEtran.bst: No hyphenation pattern has been}%
\typeout{** loaded for the language `#1'. Using the pattern for}%
\typeout{** the default language instead.}%
\else
\language=\csname l@#1\endcsname
\fi
#2}}
\providecommand{\BIBdecl}{\relax}
\BIBdecl

\bibitem{heath2016overview}
R.~W. Heath, N.~Gonzalez-Prelcic, S.~Rangan, W.~Roh, and A.~Sayeed, ``An
  overview of signal processing techniques for millimeter wave {MIMO}
  systems,'' \emph{{IEEE} J. Sel. Top. Signal Process.}, vol.~10, no.~3, pp.
  436--453, Apr. 2016.

\bibitem{JSACmmWaveMXiao2017}
M.~Xiao, S.~Mumtaz, Y.~Huang, L.~Dai, M.~Matthaiou, G.~K. Karagiannidis,
  E.~Bjornson, K.~Yang, C.~Lin, and A.~Ghosh, ``Millimeter wave communications
  for future mobile networks,'' \emph{{IEEE} J. Sel. Areas Commun.}, vol.~35,
  no.~9, pp. 1909--1935, Sep. 2017.

\bibitem{CM2019HardwareConstrainedmmWave}
X.~Yang, M.~Matthaiou, J.~Yang, C.~Wen, F.~Gao, and S.~Jin, ``Hardware
  constrained millimeter-wave systems for 5{G}: {C}hallenges, opportunities,
  and solutions,'' \emph{{IEEE} Commun. Mag.}, vol.~57, no.~1, pp. 44--50, Jan.
  2019.

\bibitem{WC2015mmWaveJYGuo}
J.~Zhang, X.~Huang, V.~Dyadyuk, and Y.~Guo, ``Massive hybrid antenna array for
  millimeter-wave cellular communications,'' \emph{{IEEE} Wireless Commun.},
  vol.~22, no.~1, pp. 79--87, Feb. 2015.

\bibitem{BOOK2007}
D.~Tse and P.~Viswanath, \emph{Fundamentals of Wireless Communication}.\hskip
  1em plus 0.5em minus 0.4em\relax Cambridge, U.K.: Cambridge Univ. Press,
  2007.

\bibitem{Access2013}
T.~S. Rappaport, S.~Sun, R.~Mayzus, H.~Zhao, Y.~Azar, K.~Wang, G.~N. Wong,
  J.~K. Schulz, M.~Samimi, and F.~Gutierrez, ``Millimeter wave mobile
  communications for 5{G} cellular: It will work!'' \emph{{IEEE} Access},
  vol.~1, pp. 335--349, 2013.

\bibitem{2010AACC}
A.~Sayeed and N.~Behdad, ``Continuous aperture phased {MIMO}: Basic theory and
  applications,'' in \emph{Proc. Allerton Conf. Commun., Control, Comput.},
  Allerton, IL, USA, 2010, pp. 1196--1203.

\bibitem{P_Schniter}
P.~Schniter and A.~Sayeed, ``{Channel estimation and precoder design for
  millimeter-wave communications: The sparse ways},'' in \emph{Proc. Asilomar
  Conf. Signals Syst. Comput.}, Pacific Grove, USA, Nov. 2014, pp. 273--277.

\bibitem{TSP2018WenyanMa}
W.~Ma and C.~Qi, ``Beamspace channel estimation for millimeter wave massive
  {MIMO} system with hybrid precoding and combining,'' \emph{{IEEE} Trans.
  Signal Process.}, vol.~66, no.~18, pp. 4839--4853, Sep. 2018.

\bibitem{RappaportTCOM2015}
T.~S. Rappaport, G.~R. MacCartney, M.~K. Samimi, and S.~Sun, ``Wideband
  millimeter-wave propagation measurements and channel models for future
  wireless communication system design,'' \emph{{IEEE} Trans. Commun.},
  vol.~63, no.~9, pp. 3029--3056, Sep. 2015.

\bibitem{TAP2014}
S.~Jaeckel, L.~Raschkowski, K.~Brner, and L.~Thiele, ``Qua{DR}i{G}a: {A} 3-{D}
  multi-cell channel model with time evolution for enabling virtual field
  trials,'' \emph{{IEEE} Trans. Antennas Propag.}, vol.~62, no.~6, pp.
  3242--3256, June 2014.

\bibitem{B_WangbeamSquintTSP2018}
B.~Wang, F.~Gao, S.~Jin, H.~Lin, and G.~Y. Li, ``{Spatial- and
  frequency-wideband effects in millimeter-wave massive MIMO systems},''
  \emph{{IEEE} Trans. Signal Process.}, vol.~66, no.~13, pp. 3393--3406, 2018.

\bibitem{B_Wang}
B.~Wang, M.~Jian, F.~Gao, G.~Y. Li, and H.~Lin, ``{Beam squint and channel
  estimation for wideband mmWave massive MIMO-OFDM systems},'' \emph{{IEEE}
  Trans. Signal Process.}, vol.~67, no.~23, pp. 5893--5908, Dec. 2019.

\bibitem{O_E_Ayach}
O.~E. Ayach, R.~W. Heath, S.~Abu-Surra, S.~Rajagopal, and Z.~Pi, ``{Low
  complexity precoding for large millimeter wave MIMO systems},'' in
  \emph{Proc. IEEE Int. Conf. Commun. (ICC)}, Ottawa, Canada, 2012, pp.
  3724--3729.

\bibitem{Access2016}
R.~M. Rial, C.~Rusu, N.~G. Prelcic, A.~Alkhateeb, and R.~W. Heath, ``Hybrid
  {MIMO} architectures for millimeter wave communications: Phase shifters or
  switches?'' \emph{{IEEE} Access}, vol.~4, pp. 247--267, 2016.

\bibitem{JSTSP2018HardwareEfficient}
X.~Yu, J.~Zhang, P.~Hwang, and K.~B. Letaief, ``A hardware-efficient analog
  network structure for hybrid precoding in millimeter wave systems,''
  \emph{{IEEE} J. Sel. Top. Signal Process.}, vol.~12, no.~2, pp. 282--297, May
  2018.

\bibitem{S_Hu}
S.~Hu, F.~Rusek, and O.~Edfors, ``{The potential of using large antenna arrays
  on intelligent surfaces},'' in \emph{Proc. {IEEE} Veh. Technol. Conf. (VTC)},
  Sydney, Australia, 2017, pp. 1--6.

\bibitem{Access2019}
E.~Basar, M.~D. Renzo, J.~D. Rosny, M.~Debbah, M.~Alouini, and R.~Zhang,
  ``Wireless communications through reconfigurable intelligent surfaces,''
  \emph{{IEEE} Access.}, vol.~7, pp. 116\,753--116\,773, 2019.

\bibitem{MDRenzoJSAC2020}
M.~D. Renzo, A.~Zappone, M.~Debbah, M.~S. Alouini, C.~Yuen, J.~d.~Rosny, and
  S.~Tretyakov, ``Smart radio environments empowered by reconfigurable
  intelligent surfaces: How it works, state of research, and the road ahead,''
  \emph{{IEEE} J. Sel. Areas Commun.}, vol.~38, no.~11, pp. 2450--2525, Nov.
  2020.

\bibitem{LDaiMDRenzoAccess2020}
L.~Dai, B.~Wang, M.~Wang, X.~Yang, J.~Tan, S.~Bi, S.~Xu, F.~Yang, Z.~Chen,
  M.~D. Renzo, C.~B. Chae, and L.~Hanzo, ``Reconfigurable intelligent
  surface-based wireless communications: Antenna design, prototyping, and
  experimental results,'' \emph{{IEEE} Access}, vol.~8, pp. 45\,913--45\,923,
  2020.

\bibitem{mmWaveVehicular2016CommMag}
J.~Choi, V.~Va, N.~Gonzalez-Prelcic, R.~Daniels, C.~R. Bhat, and R.~W. Heath,
  ``Millimeter-wave vehicular communication to support massive automotive
  sensing,'' \emph{{IEEE} Commun. Mag.}, vol.~54, no.~12, pp. 160--167, Dec.
  2016.

\bibitem{Xiao2016UAVmmWave}
Z.~Xiao, P.~Xia, and X.-G. Xia, ``Enabling {UAV} cellular with millimeter-wave
  communication: Potentials and approaches,'' \emph{{IEEE} Commun. Mag.},
  vol.~54, no.~5, 2016.

\bibitem{wang2020compressed}
P.~Wang, J.~Fang, H.~Duan, and H.~Li, ``{Compressed channel estimation for
  intelligent reflecting surface-assisted millimeter wave systems},''
  \emph{{IEEE} Signal Process. Lett.}, vol.~27, pp. 905--909, May 2020.

\bibitem{RIS2020SPAWCzhang}
J.~Zhang, C.~Qi, P.~Li, and P.~Lu, ``Channel estimation for reconfigurable
  intelligent surface aided massive {MIMO} system,'' in \emph{Proc. IEEE Int.
  Workshop Signal Process. Adv. in Wireless Commun. (SPAWC)}, Atlanta, USA, May
  2020, pp. 1--5.

\bibitem{CLiuJSAC2017}
C.~Liu, M.~Li, S.~V. Hanly, I.~B. Collings, and P.~Whiting, ``Millimeter wave
  beam alignment: Large deviations analysis and design insights,'' \emph{{IEEE}
  J. Sel. Areas Commun.}, vol.~35, no.~7, pp. 1619--1631, Aug. 2017.

\bibitem{JSAC2009}
J.~Wang, Z.~Lan, C.~Pyo, T.Baykas, C.~Sum, M.~A. Rahman, J.~Gao, R.~Funada,
  F.~Kojima, H.~Harada, and S.~Kato, ``Beam codebook based beamforming protocol
  for multi-gbps millimeter-wave {WPAN} systems,'' \emph{{IEEE} J. Sel. Areas
  Commun.}, vol.~27, no.~8, pp. 1390--1399, Oct. 2009.

\bibitem{TC2017CommonCodebook}
J.~Song, J.~Choi, and D.~J. Love, ``Common codebook millimeter wave beam
  design: Designing beams for both sounding and communication with uniform
  planar arrays,'' \emph{{IEEE} Trans. Commun.}, vol.~65, no.~4, pp.
  1859--1872, Apr. 2017.

\bibitem{XiaoCodebook2017TWC}
Z.~Xiao, P.~Xia, and X.-G. Xia, ``Codebook design for millimeter-wave channel
  estimation with hybrid precoding structure,'' \emph{{IEEE} Trans. Wireless
  Commun.}, vol.~16, no.~1, pp. 141--153, Jan. 2017.

\bibitem{K_Chen}
K.~Chen, C.~Qi, and G.~Y. Li, ``{Two-step codeword design for millimeter wave
  massive MIMO systems with quantized phase shifters},'' \emph{{IEEE} Trans.
  Signal Process.}, vol.~68, no.~1, pp. 170--180, Jan. 2020.

\bibitem{CST2019BeamManagement}
M.~Giordani, M.~Polese, A.~Roy, D.~Castor, and M.~Zorzi, ``A tutorial on beam
  management for {3GPP NR} at mmwave frequencies,'' \emph{{IEEE} Commun.
  Surveys Tuts.}, vol.~21, no.~1, pp. 173--196, 2019.

\bibitem{EMHofstetterTIT1969}
E.~M. Hofstetter and D.~Delong, ``Detection and parameter estimation in an
  amplitude-comparison monopulse radar,'' \emph{IEEE Trans. Inf. Theory},
  vol.~15, no.~1, pp. 22--30, Jan. 1969.

\bibitem{X_Sun}
X.~Sun, C.~Qi, and G.~Y. Li, ``{Beam training and allocation for multiuser
  millimeter wave massive MIMO systems},'' \emph{{IEEE} Trans. Wireless
  Commun.}, vol.~18, no.~2, pp. 1041--1053, Feb. 2019.

\bibitem{MLiTWC2019}
M.~Li, C.~Liu, S.~V. Hanly, I.~B. Collings, and P.~Whiting, ``Explore and
  eliminate: Optimized two-stage search for millimeter-wave beam alignment,''
  \emph{{IEEE} Trans. Wireless Commun.}, vol.~18, no.~9, pp. 4379--4393, Sep.
  2019.

\bibitem{DZhuTWC2017}
D.~Zhu, J.~Choi, and R.~W. Heath, ``Auxiliary beam pair enabled {A}o{D} and
  {A}o{A} estimation in closed-loop large-scale millimeter-wave {MIMO}
  systems,'' \emph{{IEEE} Trans. Wireless Commun.}, vol.~16, no.~7, pp.
  4770--4785, July 2017.

\bibitem{DZhuTWC2018}
D.~Zhu, J.~Choi, Q.~Cheng, W.~Xiao, and R.~W. Heath, ``High-resolution angle
  tracking for mobile wideband millimeter-wave systems with antenna array
  calibration,'' \emph{{IEEE} Trans. Wireless Commun.}, vol.~17, no.~11, pp.
  7173--7189, Nov. 2018.

\bibitem{S_Noh}
S.~Noh, M.~D. Zoltowski, and D.~J. Love, ``{Multi-resolution codebook and
  adaptive beamforming sequence design for millimeter wave beam alignment},''
  \emph{{IEEE} Trans. Wireless Commun.}, vol.~16, no.~9, pp. 5689--5701, Sep.
  2017.

\bibitem{Z_Xiao}
Z.~Xiao, T.~He, P.~Xia, and X.-G. Xia, ``{Hierarchical codebook design for
  beamforming training in millimeter-wave communication},'' \emph{{IEEE} Trans.
  Wireless Commun.}, vol.~15, no.~5, pp. 3380--3392, May 2016.

\bibitem{TC2-2017}
J.~Zhang, Y.~Huang, Q.~Shi, J.~Wang, and L.~Yang, ``Codebook design for beam
  alignment in millimeter wave communication systems,'' \emph{{IEEE} Trans.
  Commun.}, vol.~65, no.~11, pp. 4980--4995, Nov. 2017.

\bibitem{GYLiJSAC2017}
C.~Lin, G.~Y. Li, and L.~Wang, ``Subarray-based coordinated beamforming
  training for mm{W}ave and sub-{TH}z communications,'' \emph{IEEE J. Sel.
  Areas Commun.}, vol.~35, no.~9, pp. 2115--2126, Sep. 2017.

\bibitem{KangjianChen2021CL}
K.~Chen and C.~Qi, ``Beam training based on dynamic hierarchical codebook for
  millimeter wave massive {MIMO},'' \emph{{IEEE} Commun. Lett.}, vol.~23,
  no.~1, pp. 132--135, Jan. 2019.

\bibitem{HierarchicalBeamTraining2019Globecom}
K.~Chen, C.~Qi, O.~A. Dobre, and G.~Y. Li, ``Simultaneous multiuser beam
  training using adaptive hierarchical codebook for {mmWave} massive {MIMO},''
  in \emph{Proc. IEEE Global Commun. Conf. (GLOBECOM)}, Waikoloa, HI, USA, Dec.
  2019, pp. 1--6.

\bibitem{Chenhao2020TWCcodebook}
C.~Qi, K.~Chen, O.~Dobre, and G.~Y. Li, ``Hierarchical codebook based multiuser
  beam training for millimeter wave massive {MIMO},'' \emph{{IEEE} Trans.
  Wireless Commun.}, vol.~19, no.~12, pp. 8142--8152, Dec. 2020.

\bibitem{WenyanMaWCL2020}
W.~Ma, C.~Qi, and G.~Y. Li, ``Machine learning for beam alignment in millimeter
  wave massive {MIMO},'' \emph{IEEE Wireless Commun. Lett.}, vol.~9, no.~6, pp.
  875--878, June 2020.

\bibitem{HEchigoTVT2021}
M.~B. H.~Echigo, Y.~Cao and T.~Ohtsuki, ``A deep learning based low overhead
  beam selection in mm{W}ave communications,'' \emph{{IEEE} Trans. Veh.
  Technol.}, vol.~70, no.~1, pp. 682--691, Jan. 2021.

\bibitem{CLinVTC2019}
C.~Lin, W.~Kao, S.~Zhan, and T.~Lee, ``Bs{N}et: A deep learning based beam
  selection method for mm{W}ave communications,'' in \emph{Proc. {IEEE} Veh.
  Technol. Conf. (VTC)}, Honolulu, HI, USA, Nov. 2019, pp. 1--6.

\bibitem{KSatyTVT2019}
K.~Satyanarayana, M.~El-Hajjar, A.~A.~M. Mourad, and L.~Hanzo, ``Deep learning
  aided fingerprint-based beam alignment for mm{W}ave vehicular
  communication,'' \emph{{IEEE} Trans. Veh. Technol.}, vol.~68, no.~11, pp.
  10\,858--10\,871, Nov. 2019.

\bibitem{LocationOrientationICC2020}
S.~Rezaie, C.~N. Manchon, and E.~de~Carvalho, ``Location- and orientation-aided
  millimeter wave beam selection using deep learning,'' in \emph{Proc. IEEE
  Int. Conf. Commun. (ICC)}, Dublin, Ireland, June 2020, pp. 1--6.

\bibitem{Y_Wang}
Y.~Wang, A.~Klautau, M.~Ribero, A.~C.~K. Soong, and R.~W. Heath, ``{MmWave
  vehicular beam selection with situational awareness using machine
  learning},'' \emph{{IEEE} Access}, vol.~7, pp. 87\,479--87\,493, 2019.

\bibitem{BlockagePrediction2020TCOM}
M.~Alrabeiah and A.~Alkhateeb, ``Deep learning for {mmWave} beam and blockage
  prediction using sub-6 {GH}z channels,'' \emph{{IEEE} Trans. Commun.},
  vol.~68, no.~9, pp. 5504--5518, Sep. 2020.

\bibitem{LIDAR2019WCL}
A.~Klautau, N.~Gonzalez-Prelcic, and R.~Heath, ``{LIDAR} data for deep
  learning-based {mmWave} beam-selection,'' \emph{IEEE Wireless Commun. Lett.},
  vol.~8, no.~3, pp. 909--912, June 2019.

\bibitem{M_Hashemi}
M.~Hashemi, A.~Sabharwal, C.~Emre~Koksal, and N.~B. Shroff, ``{Efficient beam
  alignment in millimeter wave systems using contextual bandits},'' in
  \emph{Proc. IEEE Int. Conf. Comput. Commun. (INFOCOM)}, Honolulu, HI, Apr.
  2018, pp. 2393--2401.

\bibitem{VVaAccess2019}
V.~Va, T.~Shimizu, G.~Bansal, and R.~W. Heath, ``Online learning for
  position-aided millimeter wave beam training,'' \emph{{IEEE} Access}, vol.~7,
  pp. 30\,507--30\,526, Mar. 2019.

\bibitem{MChengICC2019}
M.~Cheng, J.~Wang, J.~Wang, M.~Lin, Y.~Wu, and H.~Zhu, ``A fast beam searching
  scheme in mm{W}ave communications for high-speed trains,'' in \emph{Proc.
  IEEE Int. Conf. Commun. (ICC)}, Shanghai, China, May 2019, pp. 1--6.

\bibitem{MBBoothCL2019}
M.~B. Booth, V.~Suresh, N.~Michelusi, and D.~J. Love, ``Multi-armed bandit beam
  alignment and tracking for mobile millimeter wave communications,''
  \emph{{IEEE} Commun. Lett.}, vol.~23, no.~7, pp. 1244--1248, July 2019.

\bibitem{MHussainGLOBECOM2019}
M.~Hussain and N.~Michelusi, ``Second-best beam-alignment via {Bayesian}
  multi-armed bandits,'' in \emph{Proc. IEEE Global Commun. Conf. (GLOBECOM)},
  Waikoloa, HI, USA, Dec. 2019, pp. 1--6.

\bibitem{GHSimACMTN2018}
G.~H. Sim, S.~Klos, A.~Asadi, A.~Klein, and M.~Hollick, ``An online
  context-aware machine learning algorithm for 5{G} mm{W}ave vehicular
  communications,'' \emph{{IEEE} {ACM} Trans. Networking}, vol.~26, no.~6, pp.
  2487--2500, Dec. 2018.

\bibitem{WWuTWC2019}
W.~Wu, N.~Cheng, N.~Zhang, P.~Yang, W.~Zhuang, and X.~Shen, ``Fast mm{W}ave
  beam alignment via correlated bandit learning,'' \emph{{IEEE} Trans. Wireless
  Commun.}, vol.~18, no.~12, pp. 5894--5908, Dec. 2019.

\bibitem{SKimICTC2020}
S.~Kim, G.~Kwon, and H.~Park, ``Q-learning-based low complexity beam tracking
  for mm{W}ave beamforming system,'' in \emph{Proc. Int. Conf. Inf. Commun.
  Technol. Converg. (ICTC)}, Jeju Island, Korea (South), Oct. 2020, pp.
  1451--1455.

\bibitem{JZhangGLOBECOM2019}
J.~Zhang, Y.~Huang, J.~Wang, and X.~You, ``Intelligent beam training for
  millimeter-wave communications via deep reinforcement learning,'' in
  \emph{Proc. IEEE Global Commun. Conf. (GLOBECOM)}, Waikoloa, HI, USA, Dec.
  2019, pp. 1--7.

\bibitem{A_Alkhateeb}
A.~Alkhateeb, O.~El~Ayach, G.~Leus, and R.~W. Heath, ``{Channel estimation and
  hybrid precoding for millimeter wave cellular systems},'' \emph{{IEEE} J.
  Sel. Top. Signal Process.}, vol.~8, no.~5, pp. 831--846, Oct. 2014.

\bibitem{alkhateeb2015compressed}
A.~Alkhateeb, G.~Leus, and R.~W. Heath, ``{Compressed sensing based multi-user
  millimeter wave systems: How many measurements are needed?}'' in \emph{Proc.
  IEEE Int. Conf. Acoust., Speech Signal Process. (ICASSP)}, Dubai, UAE, Apr.
  2015, pp. 2909--2913.

\bibitem{lee2016channel}
J.~Lee, G.~Gil, and Y.~H. Lee, ``{Channel estimation via orthogonal matching
  pursuit for hybrid MIMO systems in millimeter wave communications},''
  \emph{{IEEE} Trans. Commun.}, vol.~64, no.~6, pp. 2370--2386, June 2016.

\bibitem{javier2018frequency}
J.~Rodriguez-Fernandez, N.~Gonzalez-Prelcic, K.~Venugopal, and R.~W. Heath,
  ``{Frequency-domain compressive channel estimation for frequency-selective
  hybrid mmWave MIMO systems},'' \emph{{IEEE} Trans. Wireless Commun.},
  vol.~17, no.~5, pp. 2946--2960, May 2018.

\bibitem{tao2019regularlized}
J.~Tao, C.~Qi, and Y.~Huang, ``{Regularized multipath matching pursuit for
  sparse channel estimation in millimeter wave massive MIMO system},''
  \emph{{IEEE} Wireless Commun. Lett.}, vol.~8, no.~1, pp. 169--172, Feb. 2019.

\bibitem{dai2017tras}
X.~Gao, L.~Dai, S.~Han, C.-L. I, and X.~Wang, ``Reliable beamspace channel
  estimation for millimeter-wave massive {MIMO} systems with lens antenna
  array,'' \emph{{IEEE} Trans. Wireless Commun.}, vol.~16, no.~9, pp.
  6010--6021, Sep. 2017.

\bibitem{gao2016estimation}
X.~Gao, L.~Dai, S.~Han, C.-L. I, and F.~Adachi, ``{Beamspace channel estimation
  for 3D lens-based millimeter-wave massive MIMO systems},'' in \emph{Proc.
  Int. Conf. Wireless Commun. Signal Process. (WCSP)}, Yangzhou, China, Oct.
  2016, pp. 1--5.

\bibitem{ma2017channel}
W.~Ma and C.~Qi, ``{Channel estimation for 3-D lens millimeter wave massive
  MIMO system},'' \emph{{IEEE} Commun. Lett.}, vol.~21, no.~9, pp. 2045--2048,
  June 2017.

\bibitem{C_Hu}
C.~Hu, L.~Dai, T.~Mir, Z.~Gao, and J.~Fang, ``{Super-resolution channel
  estimation for mmWave massive MIMO with hybrid precoding},'' \emph{{IEEE}
  Trans. Veh. Technol.}, vol.~67, no.~9, pp. 8954--8958, Sep. 2018.

\bibitem{YingmingTsaiTCOM2018}
Y.~Tsai, L.~Zheng, and X.~Wang, ``Millimeter-wave beamformed full-dimensional
  {MIMO} channel estimation based on atomic norm minimization,'' \emph{{IEEE}
  Trans. Commun.}, vol.~66, no.~12, pp. 6150--6163, Dec. 2018.

\bibitem{anji2019channle}
C.~K. Anjinappa, Y.~Zhou, Y.~Yapici, D.~Baron, and I.~Guvenc, ``{Channel
  estimation in mmWave hybrid MIMO system via off-grid Dirichlet kernels},'' in
  \emph{Proc. IEEE Global Commun. Conf. (GLOBECOM)}, Waikoloa, HI, USA, Dec.
  2019, pp. 1--6.

\bibitem{dai2016estimation}
Z.~Gao, C.~Hu, L.~Dai, and Z.~Wang, ``{Channel estimation for millimeter-wave
  massive MIMO with hybrid precoding over frequency-selective fading
  channels},'' \emph{{IEEE} Commun. Lett.}, vol.~20, no.~6, pp. 1259--1262,
  June 2016.

\bibitem{liao20172d}
A.~Liao, Z.~Gao, Y.~Wu, H.~Wang, and M.-S. Alouini, ``{2D unitary ESPRIT based
  super-resolution channel estimation for millimeter-wave massive MIMO with
  hybrid precoding},'' \emph{{IEEE} Access}, vol.~5, pp. 24\,747--24\,757, Nov.
  2017.

\bibitem{zhang2017channel}
J.~Zhang and M.~Haardt, ``{Channel estimation and training design for hybrid
  multi-carrier mmWave massive MIMO systems: The beamspace ESPRIT approach},''
  in \emph{Proc. Eur. Signal Process. Conf. (EUSIPCO)}, Kos, Greece, Aug. 2017,
  pp. 385--389.

\bibitem{zhang2017channel2}
------, ``{Channel estimation for hybrid multi-carrier mmWave MIMO systems
  using three-dimensional unitary ESPRIT in DFT beamspace},'' in \emph{Proc.
  {IEEE} Int. Workshop Comput. Adv. Multi-Sensor Adapt. Process. (CAMSAP)},
  Curacao, Netherlands Antilles, Dec. 2017, pp. 1--5.

\bibitem{W_Ma_a}
W.~Ma, C.~Qi, and G.~Y. Li, ``{High-resolution channel estimation for
  frequency-selective mmWave massive MIMO system},'' \emph{{IEEE} Trans.
  Wireless Commun.}, vol.~19, no.~5, pp. 3517--3529, May 2020.

\bibitem{guo2017millimeter}
Z.~Guo, X.~Wang, and W.~Heng, ``{Millimeter-Wave channel estimation based on
  2-D beamspace MUSIC method},'' \emph{{IEEE} Trans. Wireless Commun.},
  vol.~16, no.~8, pp. 5384--5394, Aug. 2017.

\bibitem{wu2018long}
J.~Wu, W.~Heng, J.~Hu, X.~Li, and K.~Wang, ``{Long-term channel estimation for
  mm-Wave hybrid beamforming systems of coherent signals based on 2-D MUSIC
  algorithm},'' in \emph{Proc. {IEEE} Veh. Technol. Conf. (VTC)}, Chicago, IL,
  USA, Aug. 2018, pp. 27--30.

\bibitem{he2018deep}
H.~He, C.-K. Wen, S.~Jin, and G.~Y. Li, ``{Deep learning-based channel
  estimation for beamspace mmWave massive MIMO systems},'' \emph{{IEEE}
  Wireless Commun. Lett.}, vol.~7, no.~5, pp. 852--855, Oct. 2018.

\bibitem{wei2019amp}
Y.~Wei, M.~Zhao, M.~Zhao, M.~Lei, and Q.~Yu, ``{An AMP-based network with deep
  residual learning for mmWave beamspace channel estimation},'' \emph{{IEEE}
  Wireless Commun. Lett.}, vol.~8, no.~4, pp. 1289--1292, Aug. 2019.

\bibitem{zhang2020deep}
Y.~Zhang, Y.~Mu, Y.~Liu, T.~Zhang, and Y.~Qian, ``{Deep learning-based
  beamspace channel estimation in mmWave massive MIMO systems},'' \emph{{IEEE}
  Wireless Commun. Lett.}, vol.~9, no.~12, pp. 2212--2215, Dec. 2020.

\bibitem{wei2020deep}
X.~Wei, C.~Hu, and L.~Dai, ``{Deep learning for beamspace channel estimation in
  millimeter-wave massive MIMO systems},'' \emph{{IEEE} Trans. Commun.},
  vol.~69, no.~1, pp. 182--193, Sep. 2020.

\bibitem{jin2019channel}
Y.~Jin, J.~Zhang, S.~Jin, and B.~Ai, ``{Channel estimation for cell-free mmWave
  massive MIMO through deep learning},'' \emph{{IEEE} Trans. Veh. Technol.},
  vol.~68, no.~10, pp. 10\,325--10\,329, Oct. 2019.

\bibitem{moon2020deep}
S.~Moon, H.~Kim, and I.~Hwang, ``{Deep learning-based channel estimation and
  tracking for millimeter-wave vehicular communications},'' \emph{J. Commun.
  Netw.}, vol.~22, no.~3, pp. 177--184, June 2020.

\bibitem{W_Ma_b}
W.~Ma, C.~Qi, Z.~Zhang, and J.~Cheng, ``{Sparse channel estimation and hybrid
  precoding using deep learning for millimeter wave massive MIMO},''
  \emph{{IEEE} Trans. Commun.}, vol.~68, no.~5, pp. 2838--2849, May 2020.

\bibitem{P_Dong}
P.~Dong, H.~Zhang, G.~Y. Li, I.~Gaspar, and N.~NaderiAlizadeh, ``{Deep
  CNN-based channel estimation for mmWave massive MIMO systems},'' \emph{{IEEE}
  J. Sel. Topics Signal Process.}, vol.~13, no.~5, pp. 989--1000, Sep. 2019.

\end{thebibliography}

\end{document}